\documentclass[11pt, twoside]{scrartcl}
\usepackage{kantlipsum}
\usepackage[T1]{fontenc}
\usepackage{lmodern}
\usepackage[printonlyused]{acronym}
\usepackage{mathrsfs,bm} 
\usepackage{graphicx,cite,amssymb,amsmath}
\usepackage[format=plain,labelsep=period,font = small]{caption}
\usepackage{tikz}
\usepackage{subcaption}
\usepackage[top=3cm, bottom=3cm, left=2.5cm, right=2.5cm,tmargin=3.5cm,headsep = 1.5cm]{geometry}
\usepackage{authblk}
\usepackage{xpatch}
\usepackage{pgfplots}
\usepackage{scalerel}
\usepackage{lipsum}
\usepackage{tabu}
\usepackage[breakable, theorems, skins]{tcolorbox}
\tcbset{enhanced}
\usepackage{enumitem}
\setlist[itemize]{label={\footnotesize \textcolor{black!30}{$\blacksquare$}}, labelsep=3mm}

\usepackage{hyperref}
\hypersetup{
	unicode=false,          
	pdftoolbar=true,        
	pdfmenubar=true,        
	pdffitwindow=false,     
	pdfstartview={FitH},    
	pdftitle={Evolution of the 5G New Radio Two-Step Random Access towards 6G Unsourced MAC},    
	pdfauthor={Agostini, Chamberland, Clazzer, Dommel, Liva, Munari, Narayanan, Polaynskiy, Stanczak, Utkovski},     
	pdfsubject={Communication Theory},   
	pdfcreator={G. Liva},   
	pdfproducer={G. Liva}, 
	pdfkeywords={Random Access} {Coding} {6G}, 
	pdfnewwindow=true,      
	colorlinks=true,       
	linkcolor=black!90,          
	citecolor=red!80!black,        
	filecolor=magenta,      
	urlcolor=cyan           
}

\usepackage[]{scrlayer-scrpage}
\clearpairofpagestyles
\ohead{\small \pagemark}
\ihead{\small Evolution of the 5G NR Two-Step Random Access towards 6G UMAC}

\newtheorem{definition}{Definition}

\DeclareRobustCommand{\emphbox}[1]{%
\begin{tcolorbox}[ 
        breakable,
        left=0pt,
        right=0pt,
        top=0pt,
        bottom=0pt,
        colback=gray!15,
        colframe=black,
        width=\dimexpr\textwidth\relax, 
        enlarge left by=0mm,
        boxsep=6pt,
        arc=2pt,outer arc=2pt,
        boxrule=0.3pt,
        fontupper=\itshape,
        ]
        #1
\end{tcolorbox}
}

\newcommand{\boxtitle}[1]{{\normalfont \textbf{\textsf{#1}}}}
\newcommand{\asmp}[1]{{\normalfont\small \textsf{[A#1]}}}

\newcommand{\pilot}{\bm{x}_{\scaleto{\mathsf{P}}{3.5pt}}}
\newcommand{\pilotobs}{\bm{y}_{\scaleto{\mathsf{P}}{3.5pt}}}

\newcommand{\preamblelength}{n_{\scaleto{\mathsf{PRE}}{3.5pt}}}
\newcommand{\POlength}{n_{\scaleto{\mathsf{PO}}{3.5pt}}}

\newcommand{\PUPE}{\mathsf{PUPE}}
\newcommand{\inalpha}{\mathcal{X}}
\newcommand{\complex}{\mathbb{C}}
\newcommand{\code}{\mathcal{C}}

\newcommand{\NRX}{N_{\scaleto{\mathsf{RX}}{3.5pt}}}

\newcommand{\norm}[1]{\left\lVert#1\right\rVert}

\title{Evolution of the 5G New Radio Two-Step Random Access towards 6G Unsourced MAC
\thanks{\quad An early version of this work has been presented at the \emph{Algorithmic Structures for Uncoordinated Communications and Statistical Inference in Exceedingly Large Spaces} workshop held in Banff (Canada), March 10 - 15, 2024. P. Agostini, F. Clazzer, J. Dommel, G. Liva, A. Munari, S. Stanczak and Z. Utkovski acknowledge the financial support by the Federal Ministry of Education and Research of Germany in the program of "Souver\"an. Digital. Vernetzt." Joint project 6G-RIC, project identification numbers: 16KISK022, 16KISK020K and 16KISK030.}}
\renewcommand\footnotemark{}

\author[$\dagger$]{\normalsize Patrick Agostini}
\author[$\ddagger$]{Jean-Francois Chamberland}
\author[$\S$]{Federico Clazzer}
\author[$\dagger$]{Johannes Dommel}
\author[$\S$]{Gianluigi Liva}
\author[$\S$]{Andrea Munari}
\author[$\ddagger$]{Krishna Narayanan}
\author[$\star$]{Yury Polyanskiy}
\author[$\dagger$]{Slawomir Stanczak}
\author[$\dagger$]{Zoran Utkovski}
\affil[$\dagger$]{Fraunhofer Heinrich-Hertz-Institute}
\affil[$\ddagger$]{Department of Electrical and Computer Engineering, Texas A\&M University}
\affil[$\star$]{Laboratory for Information and Decision Systems, Massachusetts Institute of Technology}
\affil[$\S$]{German Aerospace Center, DLR}

\date{\normalsize Version 1.0 \\[3mm] \normalsize April 18, 2024}

\pgfplotsset{compat=1.17}
\begin{document}

\maketitle
\thispagestyle{empty}

\begin{abstract}
This report summarizes some considerations on possible evolutions of grant-free random access in the next generation of the 3GPP wireless cellular standard. The analysis is carried out by mapping the problem to the recently-introduced \emph{unsourced} multiple access channel (UMAC) setup. By doing so, the performance of existing solutions can be benchmarked with information-theoretic bounds, assessing the potential gains that can be achieved over legacy 3GPP schemes. The study focuses on the two-step random access (2SRA) protocol introduced by Release 16 of the 5G New Radio standard, investigating its applicability to support large MTC / IoT terminal populations in a grant-free fashion. The analysis shows that the existing 2SRA scheme may not succeed in providing energy-efficient support to large user populations. Modifications to the protocol are proposed that enable remarkable gains in both energy and spectral efficiency while retaining a strong resemblance to the legacy protocol. 
\end{abstract}

\clearpage
\tableofcontents
\thispagestyle{empty}
\clearpage

\setcounter{page}{1}
\section{Introduction}\label{sec:intro}

This report explores efficient access paradigms and algorithmic frameworks for machine-driven data transfers.
The impetus for this initiative is the increasing heterogeneity of wireless traffic, mainly driven by unattended devices.
The wireless community has recognized that traditional infrastructures designed for human-centric Internet interactions face challenges in accommodating the fundamentally different ways in which machines and \ac{IoT} devices transfer information.
Unlike humans who typically establish sustained connections and send large packets, machines transmit short sporadic payloads.
Furthermore, the density of machine is expected to far exceed that of humans in the near future.
Thus, the current architectures, based on the \emph{acquisition-estimation-scheduling} paradigm, are ill-suited for this emerging digital landscape.

To address this challenge, modern wireless standards include modalities based on efficient \ac{RA} protocols.
Such mechanisms are attuned to the unpredictable nature of machine-type communication, yet their efficiency can be enhanced.
This report explores novel access framework and algorithmic enhancements tailored for small payload communication and sporadic traffic, in the context of future 3GPP cellular standard releases.
A departure from more traditional paradigms is key to eliminate the need for individualized feedback and making high-throughput connections more cost-effective for machine-type data transfers.

In a recent release \cite{5GNR16}, the \ac{5GNR} standard included a modification of the legacy \ac{4SRA} protocol adopted by earlier releases of \ac{5GNR} as well as by the \ac{LTE} and the \ac{NB-IoT} specifications \cite{LTE,NBIoT}. The modification, aiming at reducing the latency entailed by the four-step handshake of the \ac{4SRA} protocol, reduces the \ac{RA} procedure to two steps, consisting of the transmission of an uplink message, followed by a feedback acknowledgment. The method, referred to as \ac{2SRA}, allows resuming the legacy \ac{4SRA} procedure whenever a transmission fails (details on the protocol behavior are discussed in Section \ref{sec:2SRA}). The first part of the \ac{2SRA} protocol, though, provides a first form of grant-free transmission mechanism. It is hence natural to ask whether the existing \ac{2SRA} procedure can be used as a blueprint to develop a grant-free scheme for massive \ac{MTC}/\ac{IoT} scenarios. This document aims at providing some first answers to this question. To enable a fair assessment on the efficiency of the \ac{5GNR} \ac{2SRA} protocol in massive grant-free scenarios, we use the information theoretic model of \cite{Polyanskiy2017}, referred to as \ac{UMAC} model in the following. Monte Carlo simulation results over the Gaussian \ac{MAC} and over the quasi-static fading \ac{MAC} reveal that the current specification of the \ac{2SRA} protocol --- adapted to the grant-free setting of \cite{Polyanskiy2017}--- is substantially sub-optimal in terms of both energy and spectral efficiency. Large performance gaps are identified with respect to the finite-length performance benchmarks of \cite{Polyanskiy2017}, as well as with respect to the performance achieved by recently-introduced \ac{UMAC} schemes \cite{Marshakov2019,Narayanan:SIDMA,Narayanan2020,Fengler21,Narayana20:polar}. 
The analysis of the \ac{2SRA} protocol reveals that the scheme may be improved in two main areas: by introducing an enlarged family of \ac{PRACH} preambles, and by enriching the set of patterns that users can adopt, when accessing the channel. Based on these observations, we formulate a modification of \ac{2SRA} protocol that fully embraces the use of preambles to signal a rich set of access patterns. The scheme can be viewed as a combination of two \ac{UMAC} schemes that attracted some attention in the research community, namely, the sparse \ac{IDMA} construction of \cite{Narayanan:SIDMA} and the \ac{CRDSA} / \ac{CSA} protocols of \cite{DeGaudenzi07:CRDSA,Liva11:IRSA,Liva2015:CodedAloha}. As for sparse \ac{IDMA}, the proposed protocol uses preambles to identify interleaving patterns that users will employ to transmit coded bits. As for \ac{CRDSA}/\ac{CSA}, transmissions by a user are organized in \emph{bursts} (or \emph{segments}). The receiver makes use of \ac{SIC} to mitigate multiuser interference. Its behavior  is assumed to be similar to the one introduced in \cite{Narayanan:SIDMA}, with decoding of the user data that exploits the combined observations of the segments transmitted by the user. Noting that this option is made possible by the detection of the user preambles, and that the original \ac{CRDSA}/\ac{CSA} construction relies on the decoding of the individual segments, we recognize that the proposed protocol is closer in spirit to sparse \ac{IDMA}. Leaning on this observation, and to the bursty nature of the construction, we referred to the new protocol as \ac{SB-IDMA}.

The rest of the report is organized as follows. 

\begin{itemize}
    \item Section \ref{sec:prelim} outlines the reference notation, as well as the channel models used to analyze the performance of the various schemes.
    \item Section \ref{sec:2SRA} provides a high-level description of the \ac{2SRA} protocol. Standard-specific details that are omitted/disregarded to put emphasis  on the \ac{2SRA} basic structure and performance drivers. A performance analysis over the simple models of Section  \ref{sec:prelim} is provided, focusing on configurations that are close enough to the ones used in the recent \ac{UMAC} literature. 
    \item Section \ref{sec:extensions} focuses on possible improvements of the \ac{2SRA} protocol, providing a description of \ac{SB-IDMA}. Numerical results are provided for the settings already used in the \ac{2SRA} simulations of Section \ref{sec:2SRA}.
    \item Section \ref{sec:conclusions} contains the main conclusions of the study, suggesting directions that may enable the development of a high-efficiency grant-free random access option based on \ac{2SRA} for grant-free in future releases (6G) of the 3GPP wireless cellular standard.
\end{itemize}

\bigskip

\emph{To facilitate the reading of the report, a list of the used acronyms and of the main notation elements is included at the end of the document.}

\clearpage
\section{Preliminaries}\label{sec:prelim}

\subsection{Communication Models}\label{sec:prelim:models}
In the following, two channel models will be used when comparing the performance of various random access schemes, namely the Gaussian \ac{MAC} model, and a quasi-static fading \ac{MAC} model. We denote by $\inalpha \subseteq \complex$ the alphabet used for transmission, i.e., the set of symbols that can be used by the transmitter. We assume that the user population comprises $K$ \acp{UT}, out of which $K_a\ll K$ are active. Each active user attempts the transmission of $k$ information bits. We consider transmission over $n$ channel uses, and we refer to $n$ as the \ac{UMAC} frame length. 
In the Gaussian \ac{MAC} case, the channel output (for a single channel use) is modeled as
\[
Y = \sum_{i = 1}^{K_a} X^{(i)} + Z
\]
where $X^{(i)} \in \inalpha$ is the symbol transmitted by the $i$th active \ac{UT}, and $Z$ is the complex circularly symmetric \ac{AWGN} term, i.e.,  $Z \sim \mathcal{CN}(0,\sigma^2)$. 
The sequence of symbols transmitted by the $i$th user is 
\[
\bm{X}^{(i)} = \left( X^{(i)}_1, X^{(i)}_2, \ldots, X^{(i)}_n\right)
\]
and we enforce the power constraint 
\[
||\bm{X}^{(i)}||_2^2\leq nP.
\]
The per-user \ac{SNR} is
\[
\frac{E_b}{N_0} = \frac{nP}{k\sigma^2}
\]
where $E_b$ is the energy per information bit, and $N_0$ is the single-sided noise power spectral density.
In the quasi-static fading \ac{MAC} case, we assume each \ac{UT} transmission to be affected by Rayleigh block fading, with fading coefficients that are constant over the whole \ac{UMAC} frame, and that are independent across \acp{UT}. Therefore,
\[
Y = \sum_{i = 1}^{K_a} H_i X^{(i)} + Z
\]
where $H_i \sim \mathcal{CN}(0,1)$. The average per-user \ac{SNR} is 
\[
\frac{\bar E_b}{N_0} = \frac{nP}{k\sigma^2}
\]
where $\bar E_b$ is the average energy per information bit.

Following the \ac{UMAC} model \cite{Polyanskiy2017}, all users use the same code $\mathcal{C}$ with blocklength $n$, where the number of codewords is $|\mathcal{C}|=2^k$, i.e., each user can transmit $k$ information bits. In the idealized case where the receiver is able to obtain an accurate estimate of the number of active users, the decoder outputs an unordered list $\mathcal{L}$ of $K_a$ codewords. The per-user probability of error is then given by
\begin{equation}
\PUPE = \frac{1}{K_a}\sum_{i = 1}^{K_a} \mathsf{P}\!\left(\bm{X}^{(i)} \notin \mathcal{L} \right).
\label{eq:PUPE}
\end{equation}

\emphbox{\boxtitle{Box 1: On the simulation setup.} In the simulation results provided in this report, we make use of the following assumptions:
\begin{itemize}
\item[\asmp{1}] Unless otherwise stated, pilot sequences are sampled at random, with entries that are i.i.d. complex Gaussian with zero mean and variance $P$.
\item[\asmp{2}] Similarly, if no specific statement is provided, preambles are generated at random, with entries that are i.i.d. complex Gaussian with zero mean and variance $P$. For the special case of the \ac{5GNR} two-step random access protocol, preambles are generated according to the standard, i.e., from sets of Zadoff-Chu sequences.
\item[\asmp{3}] In the definition of the signal-to-noise ratio, we will neglect the energy overhead entailed by the \ac{OFDM} \ac{CP}.
\end{itemize}
}

\clearpage
\section{5GNR Two-Step Random Access}\label{sec:2SRA}

\begin{figure}[h!]
	\centering
	\subfloat[Four-step RA\label{fig:NR-a}]{\includegraphics[width=0.3\linewidth]{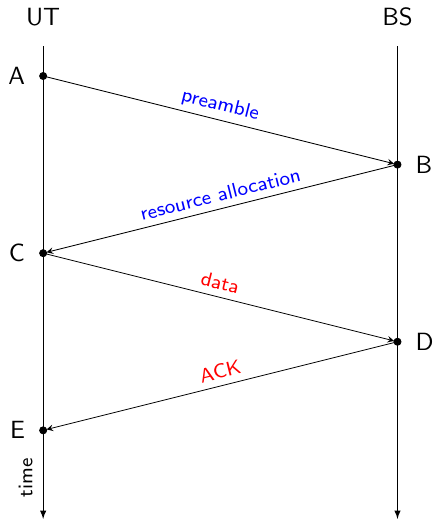}}
	\hspace{1cm}
	\subfloat[Two-step RA\label{fig:NR-b}]{\includegraphics[width=0.3\linewidth]{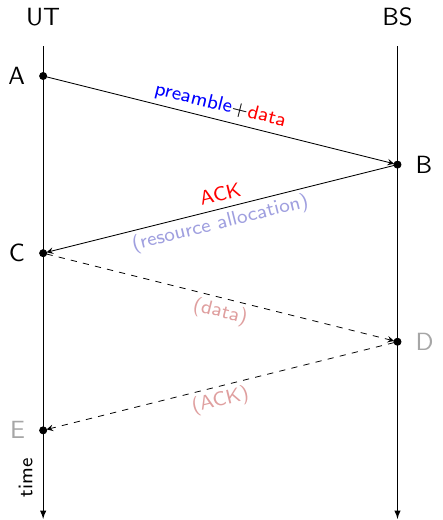}}
	\caption{Random access procedures employed by \ac{LTE}/\ac{5GNR} standards. (a) Four-step random access: the \acp{UT} transmits a preamble (\textsf{A}). Upon detecting the transmitted preambles, the \ac{BS} provides a resource allocation to each detected user (\textsf{B}). UTs transmit their data packets in the allocated resources (\textsf{C}). The BS acknowledges the correctly decoded packets (\textsf{D}). The procedure ends when the UT receives the acknowledgment (\textsf{E}). (b) Two-step random access (Release 16 of the \ac{5GNR} standard): A UT transmits a preamble, that directly points to the resource that will be used to transmit the data packet. The data packet transmission follows w/o waiting for a resource allocation (\textsf{A}). At the BS, preambles are detected and decoding is attempted in the resources pointed by the preambles. For detected UT transmissions, and acknowledgment is sent to the UTs that are successfully decoded (\textsf{B}). For detected UT transmissions that do not yield to successful decoding, orthogonal resources are allocated for the retransmission of the data packet, resuming the four-step random access procedure.}\label{fig:NR}
\end{figure}

The \ac{5GNR} random access channel inherits a four-step handshake mechanism from the \ac{LTE} standard. From a high-level point of view, the \ac{4SRA} protocol works as follows. An active user terminal picks a preamble at random for a set of (up to) $64$ Zadoff-Chu sequences. The preamble is sent over the shared \ac{PRACH}. At the base station, orthogonal resources---over a \ac{PUSCH}---are granted to each detected user preamble, with the allocation sent back to the \ac{UT}. In the subsequent phase, the \acp{UT} transmit their data units over the allocated resources. Upon decoding the transmitted packets, the receiver acknowledges the success to the transmitters. The \ac{4SRA} protocol handshake is summarized in Figure \ref{fig:NR-a}.

With Release 16 of the \ac{5GNR} standard \cite{5GNR16}, a grant-free access mechanism is introduced through the so-called \ac{2SRA} protocol. In \ac{2SRA}, each active \ac{UT} picks a random preamble from a set of (up to) $64$ Zadoff-Chu sequences, and transmits it over the \ac{PRACH}. Each preamble points to a resource, in the form of a \ac{PO}, which is used by the \ac{UT} to transmit its data unit.
The mapping can be \ac{OTO} or \ac{MTO} \cite{2step,Peralta2021}. In the \ac{OTO} mapping case, each preamble points to a distinct \ac{PO}. In the \ac{MTO} mapping case, several preambles can point to the same \ac{PO}, possibly employing different pilot sequences to enable channel estimation even in the presence of collisions.
At the \ac{BS}, the receiver  attempts demodulation/decoding at each \ac{PO} pointed by the detected preambles. For decoded packets, \ac{BS} acknowledges reception through a feedback message. If a preamble is detected, but the decoding of the associated packet transmission fails, the legacy \ac{4SRA} procedure is resumed: the \ac{BS} signals back an orthogonal resource allocation to the corresponding \ac{UT}, which proceeds with the retransmission of its packet in the granted resource unit. The procedure defined by the  \ac{2SRA} protocol is outlined in Figure \ref{fig:NR-b}.

\emphbox{\boxtitle{Box 2: Mapping the \ac{5GNR} \ac{2SRA} protocol on the \ac{UMAC} setting.} With the perspective of enabling massive grant-free connectivity, we next focus on the first phase of the \ac{2SRA}, i.e., we study its performance in isolation, excluding the possibility of exploiting feedback to resume the \ac{4SRA} grant-based procedure for unsuccessful packet transmission. We do so by removing ancillary aspects of the protocol, such as the specific mapping of the \ac{PRACH} and of \acp{PO} in the \ac{5GNR} framing structure. More specifically, we consider our channel model as a sequence of channel uses --- The reader should bear in mind that the channel uses refer to specific time/frequency resources in the \ac{5GNR} \ac{OFDM} grid. With reference to Figure \ref{fig:2step}, we denote by 
\begin{itemize}
    \item $\preamblelength$ the preamble length in \acp{c.u.};
    \item $N$ the number of available \acp{PO};
    \item $\POlength$ the length in \acp{c.u.} of each \ac{PO}.
\end{itemize}
It results that the frame length is given by 
\[
n = \preamblelength + N\times \POlength.
\]
We will consider parameters sets that may not fully be aligned with the \ac{5GNR} numerology, when seeking the need to compare \ac{2SRA} with alternative solutions. When doing so, we will nevertheless consider configurations that are sufficiently close to configurations that can be obtained using the \ac{5GNR} numerology/framing---hence, obtaining realistic estimates on the performance that can be achieved by \ac{2SRA}.}

\clearpage

\subsection{Configurations}\label{sec:2SRA:config}

In the following, we will make use of the notion of \emph{access pattern}, defined below.

\begin{figure*}[h!]
	\centering
	\includegraphics[width=0.95\textwidth]{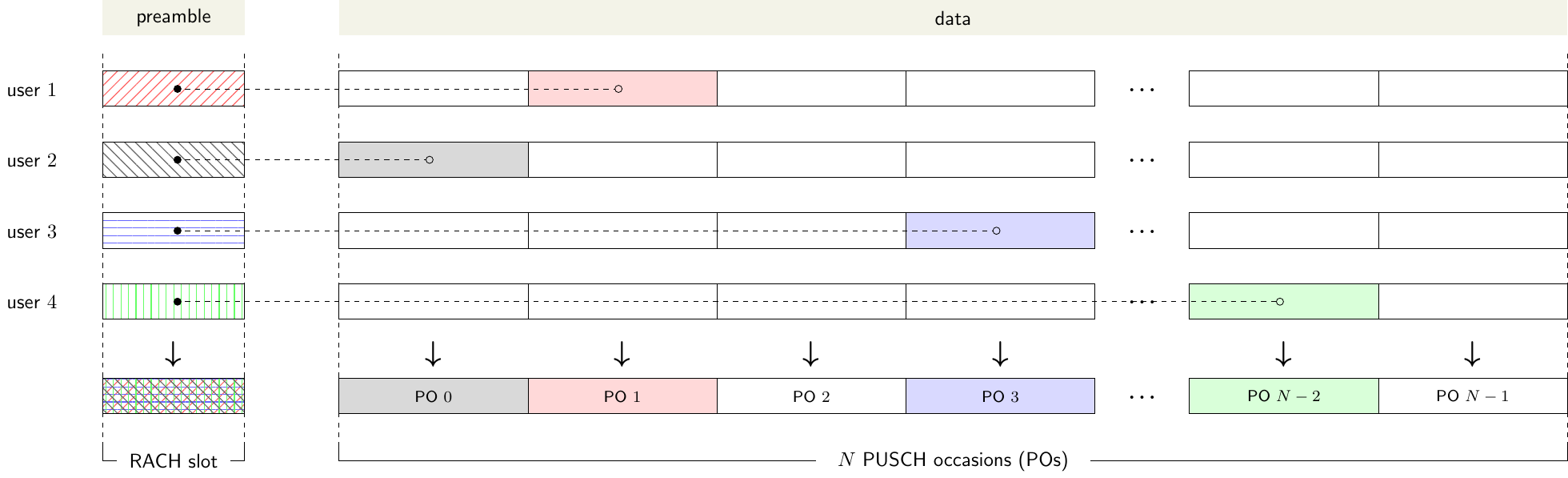}
	\caption{Illustration of the two-step random access protocol of the \ac{5GNR} standard (first transmission only). The number of \ac{PUSCH} occasions is $N$.}\label{fig:2step}
\end{figure*}

\begin{definition}[Access Pattern]\label{def:AP}
With reference to Figure \ref{fig:2step}, and assuming an order of the \acp{PO}, the access pattern of a user is defined as a binary $N$-tuple 
\[
\bm{a} = (a_0, a_1, \ldots, a_{N-1})
\]
where $a_i = 1$ if the user transmits in the $i$th \ac{PO}, and it is zero otherwise. 
\end{definition}

Access patterns define how users can access the set of available \acp{PO}.
Obviously, each preamble is associated with a specific access pattern. Figure \ref{fig:2step} provides a simplified description of the first phase of the \ac{2SRA} protocol. In the example, user $1$ selects a preamble that is associated with the access pattern with $a_1 = 1$ and $a_i = 0$ for $i\neq 1$. Upon detecting the preambles of the active users, the \ac{BS} attempts decoding at the \acp{PO} identified by the corresponding access patterns. In case of two or more users transmitting at the same \ac{PO}, correct decoding may be hindered by the mutual interference caused by the colliding users. Two configurations will be considered: 
\begin{itemize}
    \item A \ac{OTO} configuration (Table \ref{tab:2SRA}), where each preamble points to a different \ac{PO};
    \item A \ac{MTO} configuration (Table \ref{tab:2SRA_MTO}) where several preambles point to the same \ac{PO}.
\end{itemize} 
The \ac{MTO} configuration considered in the example represents an extreme case of a \ac{MTO} allocation where all preambles are mapped to a single \ac{PO}. In this case, a few random access channels are bundled together, to obtain a number of resources (channel uses) that is comparable with the \ac{OTO} configuration. Note that the parameters in the \ac{MTO} setting are provided only for the fading \ac{MAC} case, since the scheme is obviously outperformed by its \ac{OTO} counterpart over the Gaussian \ac{MAC}.

\begin{table*}[!t]
\caption{Parameters used for \ac{2SRA} simulations -- \ac{OTO} allocation.}
\centering
{\tabulinesep=1mm
\begin{tabu}{lccc}
\hline
\hline
Parameter & Gaussian \ac{MAC} & Fading \ac{MAC} & Unit\\
\hline
Information bits per user & $100$ & $100$ & bits\\
Frame length ($n$) & $16278$ & $19478$ & c.c.u.\\
\# \acp{PO} ($N$)    & $64$  & $64$ & - \\
Preamble length     & $139$ & $139$ & c.c.u.\\
Preamble repetition rate & $2$ & $2$ & - \\
Channel code & 5GNR LDPC (BG2) & 5GNR LDPC (BG2) & - \\
Block length ($n_c$) & $500$ & $500$ & bits \\
Modulation & QPSK & QPSK & - \\
\# pilot symbols per \ac{PO} & - & $50$ & - \\
\hline
\hline
\end{tabu}}
\label{tab:2SRA}
\end{table*}

\begin{table*}[!t]
\caption{Parameters used for \ac{2SRA} simulations -- \ac{MTO} allocation.}
\centering
{\tabulinesep=1mm
\begin{tabu}{lccc}
\hline
\hline
Parameter & Gaussian \ac{MAC} & Fading \ac{MAC} & Unit\\
\hline
Information bits per user & - & $100$ & bits\\
Frame length ($n$) & - & $20230$ & c.c.u.\\
\# \acp{PO} ($N$)    & -  & $1$ ($\times 35$) & - \\
Preamble length     & - & $139$ & c.c.u.\\
Preamble repetition rate & - & $2$ & - \\
Channel code & - & 5GNR LDPC (BG2) & - \\
Block length ($n_c$) & - & $500$ & bits \\
Modulation & - & QPSK & - \\
\# pilot symbols per \ac{PO} & - & $50$ & - \\
\hline
\hline
\end{tabu}}
\label{tab:2SRA_MTO}
\end{table*}

\subsection{Receiver Algorithms}

The following box provides a high-level description of the receiver algorithms used for the simulations of \ac{2SRA} (the same algorithms will be used for the simulations of the modifications of \ac{2SRA} of Section \ref{sec:extensions}).

\emphbox{\boxtitle{Box 3: On Detection and Decoding.} 
Simulations for \ac{2SRA} and for the variations introduced in Section \ref{sec:extensions} rely on the following detection / decoding algorithms. 
\begin{itemize}
    \item Preamble detection: the \ac{OMP} algorithm \cite{PatiOMP,TroppOMP} is used to detect preambles.  The algorithm is set to provide a list of $L$ preambles, where $L$ is chosen to be larger than the number of active users $K_a$. A typical value of $L$ is $\approx 1.5 K_a $. This choice allows to reduce the number of preamble misdetections. Preambles that have not been transmitted trigger a decoding stage, which is responsible of filtering out transmissions that did not take place. For each preamble identified by the \ac{OMP} algorithm, the steps below are performed.
    \item Channel estimation: on the Gaussian \ac{MAC}, the (unitary) channel coefficient of each user is assumed to be known at the receiver. On the quasi-static fading \ac{MAC}, a pilot field is included in each \ac{PO} where a user is attempting transmission. The pilot field is uniquely determined by the preamble chosen by the user. Denote by $\pilot^{(i)}$ the pilot field used by the $i$th user active in a given \ac{PO}, and by $\pilotobs$ the corresponding observation (affected by fading, \ac{AWGN} and multiuser interference). The estimate of the channel for $i$th user is obtained as 
    \begin{equation}\label{eq:CE}
    \hat{h}_i = \frac{\langle\pilotobs,\pilot^{(i)}\rangle}{\norm{\pilot^{(i)}}^2_2}.
    \end{equation}
    \item Interference-plus-Noise Power: Within each \ac{PO}, the number of active transmissions is unknown to the receiver. Hence, the power of noise plus interference needs to be estimated on a per-\ac{PO} basis. We use a blind  estimator that evaluates the interference-plus-noise power as 
    \begin{equation}\label{eq:NI}
    \mathsf{NI} = \frac{1}{\POlength}\norm{\bm{y}}^2_2
    \end{equation}
    where $\bm{y}$ is the observation associated with the \ac{PO}. 
   \item Computation of \acp{LLR}: \acp{LLR} are computed using the channel estimate provided by \eqref{eq:CE} and the interference-plus-noise power estimate from \eqref{eq:NI}. Note that, although the quasi-static fading \ac{MAC} maintains the channel coefficients of the active users constant across a frame, channel estimation will be performed on a per-\ac{PO} basis. Hence, the \acp{LLR} computed within a \ac{PO} will be based only on the \ac{PO}-specific channel estimate. Moreover, we shall observe that the interference-plus-noise power estimate from \eqref{eq:NI} does not depurate the estimate from the useful signal power. Therefore, it tends to provide a pessimistic quantification of the interference-plus-noise power.
   \item Decoding: When packets are encoded with \ac{LDPC} codes, we assume \ac{BP} decoding with $50$ iterations. When packets are encoded with polar codes, we assume \ac{SCL} decoding with adaptive list size, with maximum list size set to $128$ \cite{TV15,Li2012}.
   \item Packet Validation: For \ac{LDPC} codes, error detection relies on the output of the \ac{BP} decoder: an error flag is raised in the decoded word does not fulfill the code parity-check equations. For the polar code case, a specific error detection method will be discussed in some detail in Section \ref{sec:extensions}.
\end{itemize}
}

A receiver that operates according to the steps described above is referred to as \ac{TIN} receiver. We will consider also the improvement obtained by applying \ac{SIC}, where
\begin{itemize}
    \item When a packet is correctly decoded, the corresponding modulated codeword is used, together with the pilot field, to re-estimate the channel coefficient using an approach that is analogous to the one of \eqref{eq:CE}. Using the improved channel  estimate, the pilot field and codeword contributions are removed (cancelled) from the corresponding \ac{PO}.
    \item The contribution of the corresponding preamble is also subtracted from the \ac{PRACH}. The channel estimate used for the preamble cancellation is the one delivered by the \ac{OMP} algorithm.
    \item After interference cancellation, the whole procedure described in Box 3 is performed again, i.e., preamble detection followed by the decoding attempts over the residual signal. The iterative procedure is repeated until no more valid packets are recovered.
\end{itemize}

Note that the estimation / detection procedures outlined above can be improved, for example, by developing more sophisticated interference-plus-noise power estimators, by performing joint decoding and channel estimation, by adopting more powerful preamble detection techniques, and (in case of non-orthogonal pilot sequences) by replacing \eqref{eq:CE} with a \ac{MMSE} channel estimator. For the results provided in this document, it was decided to use simpler --- yet, suboptimal --- algorithms which may well reflect the behavior of existing implementations. 

\subsection{Two-Step Random Access: Gaussian MAC Performance}

 To enable comparisons with recently-proposed \ac{UMAC} schemes, the protocol parameters have been customized to approximate a classical scenario addressed in the \ac{UMAC} literature, which relies on a \ac{MAC} frame composed by $15000$ \acp{c.c.u.} --- equivalent to $30000$ real \acp{c.u.}. The parameters used for the simulation refer to the \ac{OTO} configuration, which achieves the best performance over the Gaussian \ac{MAC}, and are summarized in Table \ref{tab:2SRA}. In a nutshell, the configuration relies on 
\begin{itemize}
    \item A frame length of $16278$ \acp{c.c.u.}, equivalent to $32556$ real \acp{c.u.};
    \item $64$ \acp{PO}, where each \ac{PO} consists of $250$ \acp{c.c.u.};
    \item Within a \ac{PO}, transmission takes place by means of the rate-$1/5$ \ac{5GNR} $(500,100)$ \ac{LDPC} code (derived from the so-called \emph{base graph} $2$ \cite{Ric18});
    \item \ac{QPSK} modulation;
    \item Preambles chosen from the Zadoff-Chu dictionary with length $139$, with two-fold preamble repetition according to configuration A1 \cite[Chapter 16]{DAHLMAN2018}.
\end{itemize}
Figure \ref{fig:AWGN} reports the performance of \ac{2SRA} over the Gaussian \ac{MAC}.
The performance is provided in terms of minimum \ac{SNR} required to achieve a target \ac{PUPE} of $5\times 10^{-2}$. On the same chart the performance of several \ac{UMAC} schemes from the literature is shown, together with the \ac{UMAC} achievability \ac{RCU} bound of \cite{Polyanskiy2017} (\ref{fig:AWGN:RCU}). The bound as well as the competing schemes assume $30000$ real \acp{c.u.}. The schemes for which the performance is provided are $T$-fold \ac{IRSA} \cite{Marshakov2019} (\ref{fig:AWGN:TIRSA}), sparse \ac{IDMA} \cite{Narayanan:SIDMA} (\ref{fig:AWGN:SIDMA}), \ac{CCS} \cite{Narayanan2020} (\ref{fig:AWGN:CCS}), the \ac{CCS} \ac{SPARC}-based construction of \cite{Fengler21} (\ref{fig:AWGN:SPARC}), and the spread-spectrum scheme with data-dependent spreading codes from \cite{Narayana20:polar} (\ref{fig:AWGN:SPREAD}). 

The performance for the tested \ac{OTO} \ac{2SRA} configuration shows a remarkable gap from the Gaussian \ac{MAC} achievability bound, as well as from the reference \ac{UMAC} scheme from literature. The gap is visible under both \ac{TIN} (\ref{fig:AWGN:2SRA_TIN}) and \ac{TIN}-\ac{SIC} (\ref{fig:AWGN:2SRA_TINSIC}) decoding, with a saturation of the number of supported users at high \ac{SNR} that is around $K_a = 15$ in the former case, and $K_a = 35$ in the latter case.

\subsection{Two-Step Random Access: Quasi-Static Fading MAC Performance}

Over the quasi-static fading \ac{MAC}, the performance is provided only for the case of \ac{TIN}-\ac{SIC} decoding. Figure \ref{fig:fading_SISO1} depicts the performance in terms of minimum \ac{SNR} required to achieve a target \ac{PUPE} of $10^{-1}$. The results are given for a frame length in the order of $20000$ \acp{c.c.u.}. As theoretical reference, a non-rigorous estimation on the performance of the optimal decoder applied to a Gaussian codebook (via replica method) \cite{Frolov20} is provided (\ref{fig:SISO1:ASYMPTOTIC}). The estimation is developed in the asymptotic regime $n\rightarrow \infty$, and adapted to the finite frame length as suggested in \cite{Frolov20}.

Both the \ac{OTO} (\ref{fig:SISO1:2SRA_OTO}) and the \ac{MTO} (\ref{fig:SISO1:2SRA_MTO}) configuration yield a performance that is far from the reference. The \ac{MTO} allocation shows a largely improved performance over the \ac{OTO} allocation. The reason for the gain stems from the improved \ac{MPR} provided by the \ac{MTO} mapping: while the average number of users colliding in a \ac{PO} is larger than in the \ac{OTO} case (due to the smaller number $N$ of available \acp{PO}, see Tables \ref{tab:2SRA} and \ref{tab:2SRA_MTO}), most of the times colliding users employ a different preamble and, hence, a different pilot sequence, allowing for a sufficiently-accurate estimate of the corresponding channel coefficients. On the contrary, in the \ac{OTO} allocation, two users colliding in the same \ac{PO} use the same preamble, hence, the same pilot sequence --- hindering the possibility of accurate channel estimation when two users have similar channel gains.

\subsection{Two-Step Random Access: Some Considerations}\label{sec:2SRA:Considerations}

\begin{figure}
    \centering
    \begin{tikzpicture}
	\pgfplotsset{set layers,compat=1.3}
	
	\begin{axis}[%
		width=0.9\textwidth,
		height=0.7\textwidth,
		scale only axis,
		ymin=0,
		ymax=180,
		ylabel={Number of active users $K_a$},
		xmin=7,
		xmax=20,
		xlabel={$\bar{E}_b/N_0$ [dB]},
		xlabel near ticks,
		grid=both,
		grid style={dashed,gray!50},
		minor grid style={dotted,gray!40},
		minor tick num=3,
		legend cell align={left},
		legend style={font= \footnotesize, draw = none},
		]

		\addplot [color=red,mark=*,line width=0.5pt,dashed,mark options={solid,fill=white}]
	table[y=K,x=snr] {charts/TwoStep500LDPC_fading_2xPre.txt}; \label{fig:SISO1:2SRA_OTO}
	
		\addplot [color=red,mark=square*,line width=0.5pt,dashed,mark options={solid,fill=white}]
	table[y=K,x=snr] {charts/TwoStep500LDPC_fading_1024preambles_2xPre.txt};
    \label{fig:SISO1:2SRA_OTO_1024}
 
	\addplot [color=red,mark=triangle*,line width=0.5pt,dashed,mark options={solid,fill=white}]
	table[y=K,x=snr] {charts/TwoStep500LDPC_fading_8192preambles_2xPre_idealdet.txt}; \label{fig:SISO1:2SRA_OTO_8192}
	
	\addplot [color=red,mark=diamond*,line width=0.5pt,dashed,mark options={solid,fill=white}]
	table[y=K,x=snr] {charts/TwoStep500LDPC_fading_16384preambles_2xPre_idealdet.txt};\label{fig:SISO1:2SRA_OTO_16384}
	
		\addplot [color=blue,mark=*,line width=0.5pt,dashed,mark options={solid,fill=white}]
	table[y=K,x=snr] {charts/TwoStep_Fading_Alternate_Config.txt};\label{fig:SISO1:2SRA_MTO}
	
	\node[align=center,scale = 0.8] at (axis cs: 18.4,32) {\footnotesize \textsf{5GNR two-step RA} \\[-1mm] \footnotesize \textsf{OTO, $64$ preambles}};
	\draw[-latex,line width = 0.5pt] (axis cs: 18.2,28) -- (axis cs: 18.3,21);
	
	\node[align=center,scale = 0.8] at (axis cs: 18.3,65) {\footnotesize \textsf{5GNR two-step RA} \\[-1mm] \footnotesize \textsf{MTO, $64$ preambles}};
	\draw[-latex,line width = 0.5pt] (axis cs: 18.1,60) -- (axis cs: 18.3,53);
	
	\node[align=center,scale = 0.8] at (axis cs: 15.8,103) {\footnotesize \textsf{OTO} \\[-1mm] \footnotesize \textsf{$1024$ preambles}};
	\draw[-latex,line width = 0.5pt] (axis cs: 16.5,97) -- (axis cs: 17.3,77);
	
	\node[align=center,scale = 0.8] at (axis cs: 17,126) {\footnotesize \textsf{OTO}\\[-1mm] \footnotesize \textsf{$8192$ preambles}};
	\draw[-latex,line width = 0.5pt] (axis cs: 17.2,120) -- (axis cs: 18,93);
	
	\node[align=center,scale = 0.8] at (axis cs: 18.3,145) {\footnotesize \textsf{OTO}\\[-1mm] \footnotesize \textsf{$16384$ preambles}};
	\draw[-latex,line width = 0.5pt] (axis cs: 18.3,138) -- (axis cs: 18.5,103);
	
	\addplot [color=black,line width=0.5pt]
	table[y=K,x=snr] {charts/fading_replica.txt};
	
	\node[scale = 0.9,align = center,color = black, rotate  = 88.5] at (axis cs: 8.0,100) {\footnotesize \textsf{Optimum Decoding} \\ \footnotesize \textsf{(Asymptotic, Replica Method) \cite{Frolov20}} }; \label{fig:SISO1:ASYMPTOTIC}
		
	\end{axis}
\end{tikzpicture}
    \caption{Number of supported active users vs. average \ac{SNR} for a target $\PUPE = 10^{-1}$. Quasi-static Rayleigh fading channel, $n\approx 20000$ channel uses. Single antenna at the base station. The \ac{2SRA} performance is provided for the \ac{OTO} configuration (Table \ref{tab:2SRA}).}
    \label{fig:fading_SISO1}
\end{figure}
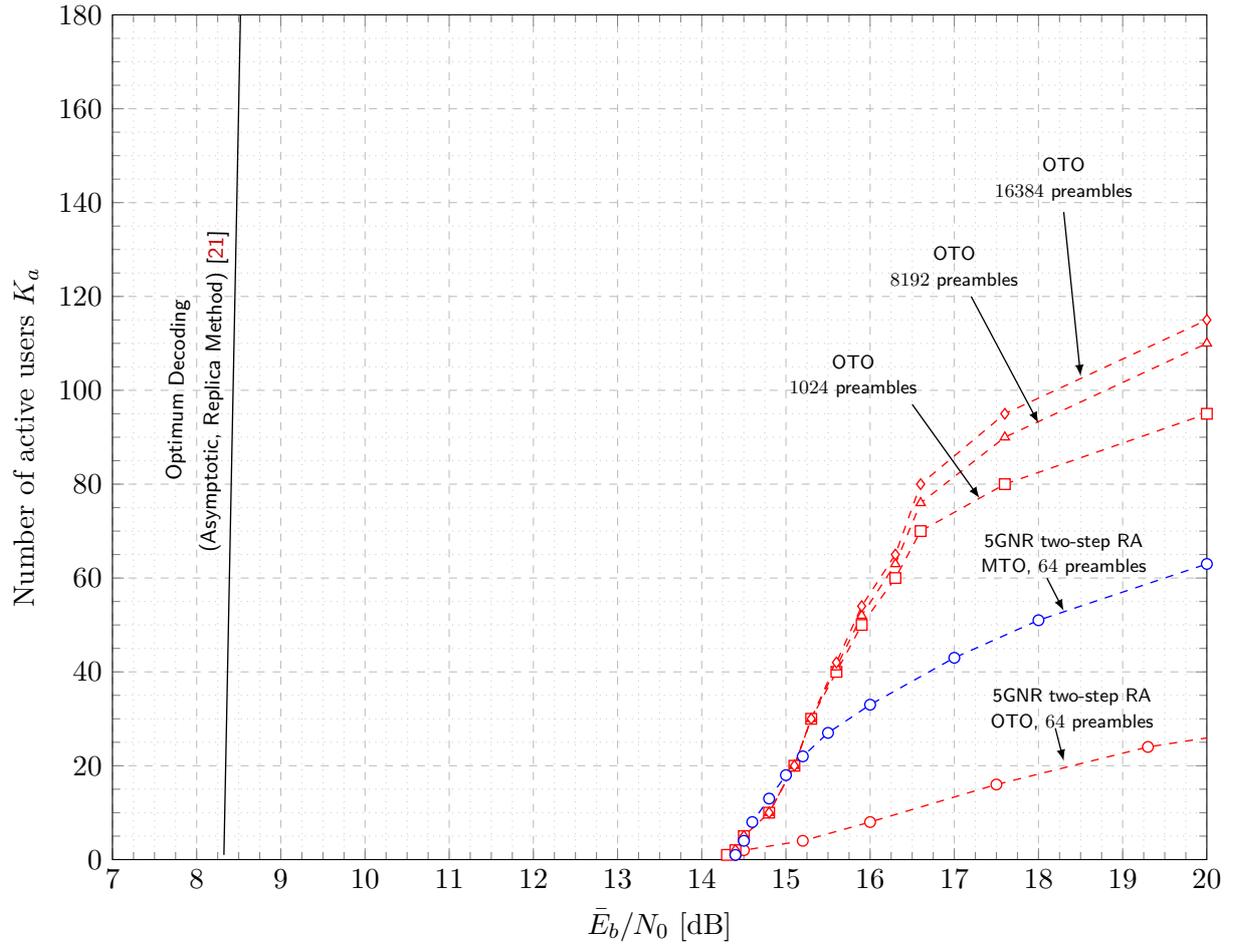

The performance of the tested \ac{2SRA} configurations points to the following observations:
\begin{itemize}
\item \ac{MTO} configurations allow to leverage from an improved \ac{MPR} capability, with respect to the \ac{OTO} case.
\item This comes at the expense on the number of available \acp{PO}: since more \acp{PRACH} have to be allocated, a smaller number of \acp{PO} can be allocated for a fixed total number of channel uses.
\item The tension between the two points above may be solved by increasing the preamble set size, enabling the use of \ac{OTO} mappings with a single \ac{PRACH}, and the larger number of \acp{PO}. This point will be at the core of the solutions investigated in Section \ref{sec:extensions}.
\item There are other two main limitations of the existing \ac{2SRA} design, which may not be immediately recognized from the simulation results in Figure \ref{fig:AWGN} and in Figure \ref{fig:fading_SISO1}: the limited capability of the \ac{5GNR} \ac{LDPC} codes to resolve multiple collisions, and the inherent lack of diversity with respect to interference. The first point is due to the lowest rate achievable by the \ac{5GNR} \ac{LDPC} codes, which is $1/5$. Lower rates may allow for a stronger \ac{MPR} capability. This issue is exacerbated, at small block length, by the tangible suboptimality of short \ac{LDPC} codes. The second point is due to the limited number of access patterns entailed by the \ac{5GNR} \ac{2SRA} scheme: the number of access patterns is given by $N$, i.e., by the number of \acp{PO}. The combined effect of these two points is to further limit the performance achievable by the \ac{5GNR} \ac{2SRA} protocol.
\item \ac{2SRA} implements a variation of slotted Aloha \cite{Abramson:ALOHA,Roberts1975}, where the \acp{PO} selected by the users are announced by a preamble transmitted over the \ac{PRACH}. One may observe that slotted Aloha does not require additional preamble transmission: a classical slotted Aloha receiver would simply attempt decoding at every available \ac{PO}. By omitting the transmission of the preamble, a significant energy savings may be achieved, shifting the \ac{2SRA} curves of Figures \ref{fig:AWGN} and \ref{fig:fading_SISO1} left by approx. 
\[
10 \log_{10} \frac{\preamblelength + \POlength}{\POlength} = 2.84 \,\mathrm{[dB]}.
\]
A natural question would be then if \ac{2SRA} could be operated without the initial preamble transmission. System-level considerations lead us to exclude this possibility, since the initial preamble transmission serves additional purposes. As discussed earlier, it allows to resume the \ac{4SRA} procedure in case of decoding failure. Moreover, preambles allow measuring the delay of user transmissions, providing essential information to implement \ac{TA} \cite[Chapter 15]{DAHLMAN2018}. Therefore, the possibility of sparing the preamble transmission should be carefully addressed taking into account its impact at the system level.
\end{itemize}

\clearpage
\begin{figure}
    \centering
    \begin{tikzpicture}
	\pgfplotsset{set layers,compat=1.3}
	
	\begin{axis}[%
		width=0.9\textwidth,
		height=0.7\textwidth,
		scale only axis,
		ymin=0,
		ymax=150,
		ylabel={Number of active users $K_a$},
		xmin=-1,
		xmax=10,
		xlabel={$E_b/N_0$ [dB]},
		xlabel near ticks,
		grid=both,
		grid style={dashed,gray!50},
		minor grid style={dotted,gray!40},
		minor tick num=3,
		legend cell align={left},
		legend style={font= \footnotesize, draw = none},
		]

		\addplot [color=black,line width=0.5pt, forget plot]
		table[y=K,x=snr] {charts/RCU.txt}; \label{fig:AWGN:RCU}

		\addplot [color=blue,line width=0.5pt,mark=x,mark options={solid}]
		table[y=K,x=snr] {charts/IRSApolar.txt};
		\addlegendentry{$T$-fold IRSA \cite{Marshakov2019}}; \label{fig:AWGN:TIRSA}

		\addplot [color=blue!50!yellow,solid,line width=0.5pt,mark=square*,mark options={solid},mark size=1.5pt]
		table[y=K,x=snr] {charts/IDMA.txt};
		\addlegendentry{Sparse IDMA \cite{Narayanan:SIDMA}}; \label{fig:AWGN:SIDMA}
		
		\addplot [color=blue!50!red,solid,line width=0.5pt,mark=triangle*,mark options={solid,rotate=90}]
		table[y=K,x=snr] {charts/CCS.txt};\addlegendentry{CCS \cite{Narayanan2020}}; \label{fig:AWGN:CCS}
		
		\addplot [color=red,solid,line width=0.5pt,mark=+,mark options={solid}]
		table[y=K,x=snr] {charts/SPARCS.txt};
		\addlegendentry{CCS with SPARCs \cite{Fengler21}}; \label{fig:AWGN:SPARC}

		\addplot [color=red!50!black,solid,mark=*,line width=0.5pt]
		table[y=K,x=snr] {charts/SpreadPolar.txt};
		\addlegendentry{Spread spectrum (polar codes) \cite{Narayana20:polar}}; \label{fig:AWGN:SPREAD}

		\addplot [color=green!50!black,solid,line width=0.5pt,mark=square*,mark options={solid,fill=white},mark size=2pt]
		table[y=K,x=snr] {charts/BSIDMAd4PC1000.txt};\addlegendentry{SB-IDMA, $(1000,100)$ polar code}; \label{fig:AWGN:SBIDMA_POLAR}
		
		\addplot [color=blue!60,solid,line width=0.5pt,mark=*,mark options={solid,fill = white}]
		table[y=K,x=snr] {charts/BSIDMAd8LDPC500.txt};\addlegendentry{SB-IDMA, $(500,100)$ LDPC code}; \label{fig:AWGN:SBIDMA_LDPC}

		\addplot [color=red,dashed,mark=*,line width=0.5pt,mark options={solid,fill=white}]
		table[y=K,x=snr] {charts/TwoStep500LDPC.txt}; \label{fig:AWGN:2SRA_TINSIC}
		
		\addplot [color=red,dashed,mark=star,line width=0.5pt,mark options={solid,fill=white}]
		table[y=K,x=snr] {charts/TwoStep500LDPC_noSIC.txt}; \label{fig:AWGN:2SRA_TIN}
		
		\node at (axis cs: -0.5,138) {\footnotesize \textsf{RCU} \cite{Polyanskiy2017}};
		\draw[-latex,line width = 0.5pt] (axis cs: -0.3,135) -- (axis cs: 0.4,130);
		
		\node[align=center] at (axis cs: 6,45) {\footnotesize \textsf{Two-Step Random Access} \\ \footnotesize $(500,100)$ \textsf{LDPC Code}};
		\draw[-latex,line width = 0.5pt] (axis cs: 6.2,40) -- (axis cs: 7.2,15);
		\draw[-latex,line width = 0.5pt] (axis cs: 6.2,40) -- (axis cs: 6.8,9.5);
		\node at (axis cs: 8.1,26) {\footnotesize \textsf{TIN-SIC}};
		\node at (axis cs: 9,9) {\footnotesize \textsf{TIN}};
		
	\end{axis}
\end{tikzpicture}
    \caption{Number of supported active users vs. \ac{SNR} for a target $\PUPE = 5 \times 10^{-2}$. AWGN channel, $n \approx 15000$ \acp{c.c.u.}.}
    \label{fig:AWGN}
\end{figure}
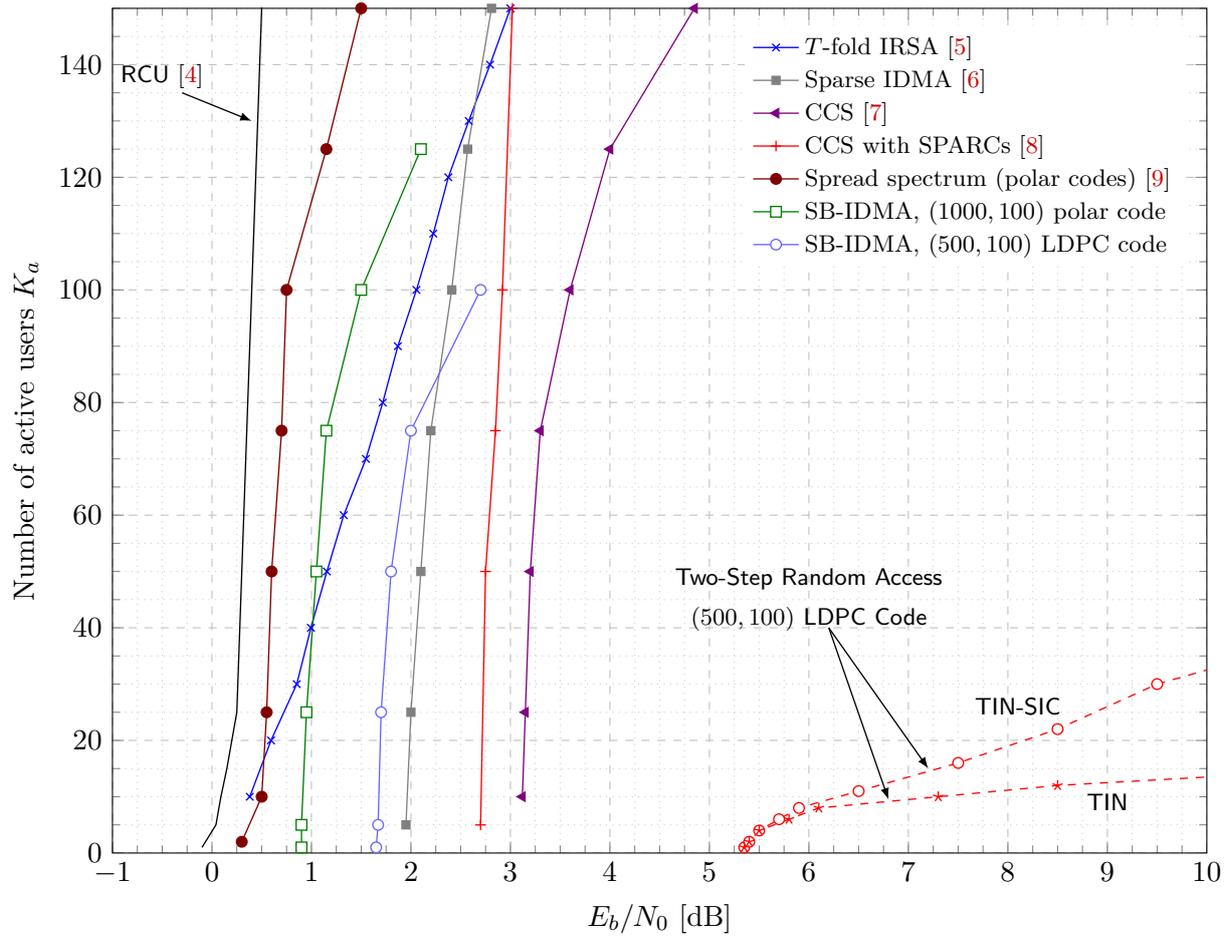

\clearpage
\section{Improving the 5G NR Two-Step Random Access}\label{sec:extensions}

This section investigates possible improvements of the \ac{5GNR} \ac{2SRA} protocol. The modifications that will be outlined in the following subsections aim at addressing some of the issues identified in Section \ref{sec:2SRA:Considerations}, namely:
\begin{itemize}
\item We study the impact of using larger preamble sets (Section \ref{sec:improvements:morepreambles}).
\item Inspired by the sparse \ac{IDMA} scheme of \cite{Narayanan:SIDMA}, we introduce a richer access pattern family, that adapts sparse \ac{IDMA} to the framing structure of \ac{5GNR}. We refer to the proposed scheme as \ac{SB-IDMA} (Section \ref{sec:improvements:SBIDMA}).
\item For \ac{SB-IDMA}, we analyze the performance gain that can be achieved, in the short block length regime, by replacing the \ac{5GNR} \ac{LDPC} codes \cite{Ric18} with the \ac{5GNR} polar codes \cite{Bio20}.
\end{itemize}

\subsection{Extended Preamble Set}\label{sec:improvements:morepreambles}

A first improvement over the \ac{5GNR} \ac{2SRA} protocol may be obtained by using preamble sets with cardinality larger than the one of the standardized Zadoff-Chu sequence family. The rationale behind this choice is to remove the tension between the limited \ac{MPR} capability of \ac{OTO} configurations (due to the use of identical pilot sequences for colliding users) and the preamble transmission overhead incurred by a repeated use of \ac{MTO} configurations (see Section \ref{sec:2SRA:Considerations}).

Figure \ref{fig:fading_SISO1} reports the performance achievable over the quasi-static fading \ac{MAC} by enlarging the preamble set to $1024$ (\ref{fig:SISO1:2SRA_OTO_1024}), $8192$ (\ref{fig:SISO1:2SRA_OTO_8192}), and $16384$ (\ref{fig:SISO1:2SRA_OTO_16384}) preambles. In all three cases, non-orthogonal preambles with symbols drawn independently from a complex Gaussian distribution were used. The preamble length has been fixed to the one used in the \ac{2SRA} simulations (see Section \ref{sec:2SRA:config}), i.e., $\preamblelength = 2 \times 139$ \acp{c.c.u.}. Each preamble points one of the $N=64$ \acp{PO}, realizing a \ac{MTO} configuration, and to a unique pilot sequence. The rest of the configuration parameters is unmodified with respect to Table \ref{tab:2SRA}. The result with $1024$ preambles shows a remarkable gain over the \ac{OTO} mapping configuration with $64$ preambles: the number of active users supported by the system is multiplied by a factor four. The improvement over the \ac{MTO} configuration with $64$ preambles is remarkable, too, although limited to a $50\%$ gain in the number of supported users. 

The performance for the case of $8192$ and $16384$ assumes ideal preamble detection. The reason for this is that, when performing preamble detection via \ac{OMP}, a large preamble misdetection rate yields a visible performance degradation, calling for the use of stronger preamble detection algorithms, or for an optimized preamble set design. Nevertheless, the results provide still an important insight, that is, enlarging the preamble set beyond a certain limit gives diminishing returns. In fact, the performance achieved with $8192$ under ideal preamble detection only marginally improves the performance obtained by the configuration employing $1024$ preambles (and actual, \ac{OMP}-based, preamble detection).

Even for $16384$ preamble with ideal detection, the performance shows signs of saturation at about $100\div 120$ supported active users. The reason of this lies mainly in the limited \ac{MPR} capability provided by the error correction scheme, and by the limited number of access patterns inherent to the slotted Aloha nature of \ac{2SRA} (see Section \ref{sec:2SRA:Considerations}). These points are addressed in the following subsection.

\subsection{Sparse-Block Interleaver Diversity Multiple Access}\label{sec:improvements:SBIDMA}

The modification of the \ac{2SRA} proposed in this section is inspired by the sparse \ac{IDMA} scheme of \cite{Narayanan2020}. In a nutshell, sparse \ac{IDMA} works as follows:
\begin{itemize}
\item Each user splits its message in two parts: a first part, that is used to selected a preamble, as well as an interleaving pattern, and a second part that is encoded via a binary linear block code, generating a codeword $\bm{c}$.
\item The codeword $\bm{c}$ is repeated $d$ times. Zero-padding follows up, resulting in a vector $\bm{v}$ of a prescribed vector length.
\item The preamble is transmitted over the channel, followed by an interleaved version of $\bm{v}$, with interleaver determined by the first message part (and, hence, by the preamble).
\item At the receiver side, upon detecting the preambles of the active users, joint decoding of the users transmission is performed. In particular, in \cite{Narayanan2020} the channel code used to encode the second message part is a suitably-designed \ac{LDPC} code. Joint decoding is performed by exploiting the knowledge of the users interleaving patterns obtained from the detected preambles: \ac{BP} decoding is performed over the joint Tanner graph \cite{Tan81} of the detected users. 
\end{itemize}
The \ac{SB-IDMA} approach closely resembles these steps, with some important differences. First, the entire user message is encoded via an $(n_c,k)$ binary linear block code
, no splitting of the user message is performed. Still, the user message is used to select the preamble by hashing the message. Second,  while the interleaving used in sparse \ac{IDMA} applies an unconstrained permutation at the symbol level, in \ac{SB-IDMA} the interleaver is restricted to permute blocks of symbols, where each block (segment, in the following) is allocated to specific \ac{PO}. In other words, the interleaver used in \ac{SB-IDMA} defines the user access pattern (see Definition \ref{def:AP}). This choice stems from the need of adapting sparse \ac{IDMA} to the \ac{5GNR} framing, and to transmission over fading channels. The key intuition is that, by clustering symbols in segments, channel estimation can be performed on a per-\ac{PO} basis by appending a pilot field to each segment. The use of sufficiently-large segments allows to reduce the pilot field overhead. Finally, at the receiver side, we do not assume joint multi-user decoding. We rather consider the use of single-user detection and decoding, enhanced by \ac{SIC} (\ac{TIN}-\ac{SIC}).

\subsubsection{Transmission Chain}

\begin{figure}
    \centering
    \includegraphics[width = 0.9\textwidth]{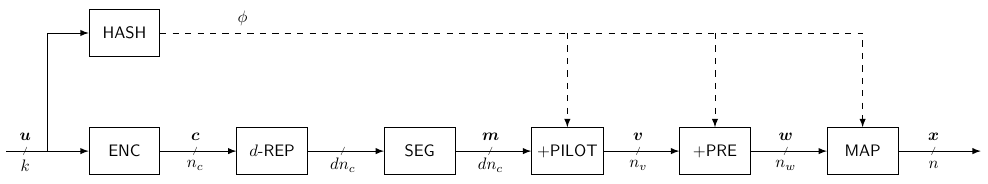}
    \caption{\ac{SB-IDMA} transmission chain.}
    \label{fig:SBIDMA_ENC}
\end{figure}

\begin{figure}
    \centering
    \includegraphics[width = \textwidth]{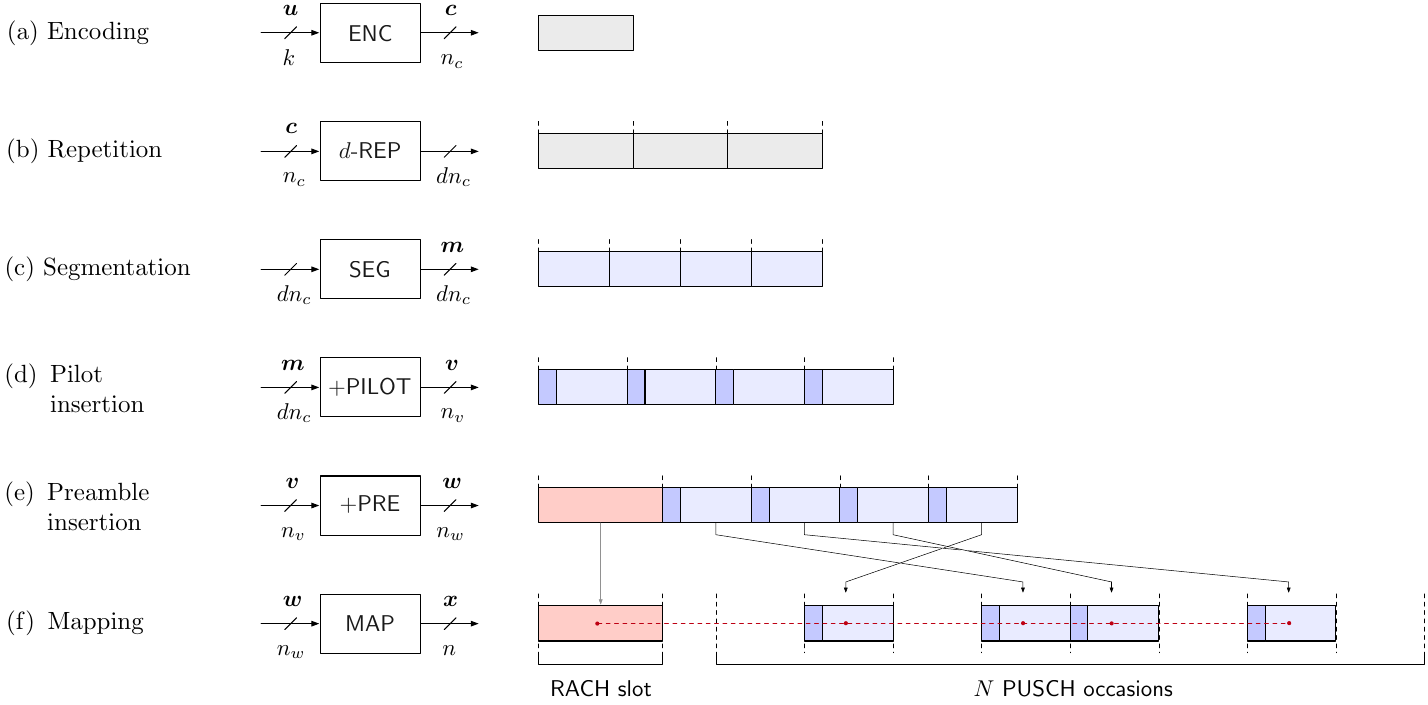}
    \caption{Detailed description of the transmission steps of Figure \ref{fig:SBIDMA_ENC}.}
    \label{fig:SBIDMA_detail}
\end{figure}

The transmission diagram of \ac{SB-IDMA} is given in Figure \ref{fig:SBIDMA_ENC}, with steps that are detailed in Figure \ref{fig:SBIDMA_detail}. Recalling that the access frame is composed by $N$ \acp{PO}, the scheme works as follows:
\begin{itemize}
\item[1.] The $k$-bits message $\bm{u}$ is hashed, generating an index $\phi(\bm{u})$ which uniquely identifies (a) a preamble within the preamble dictionary, (b) a set of $n_s$ pilot sequences, and a set of $n_s$ distinct indexes in $[1,2,\ldots,N]$.
\item[2.] The message $\bm{u}$ is encoded via a binary linear block code $\code$. The resulting codeword is \ac{QPSK}-modulated, resulting in the $n_c$-symbols vector $\bm{c}$ (Figure \ref{fig:SBIDMA_detail}(a)).
\item[3.] The modulated codeword $\bm{c}$ is repeated $d$ times (Figure \ref{fig:SBIDMA_detail}(b)).
\item[4.] The resulting vector, composed by $dn_c$ \ac{QPSK} symbols, is split into $n_s$ segments of equal length (Figure \ref{fig:SBIDMA_detail}(c)).
\item[5.] Each of the $n_s$ pilot fields identified in step 1 is appended to the respective segment, i.e., the first pilot field is appended to the first segment, the second pilot field is appended to the second segment, etc.  (Figure \ref{fig:SBIDMA_detail}(d)).
\item[6.] The preamble selected in Step $1$ is appended to the sequence of segments (Figure \ref{fig:SBIDMA_detail}(e)).
\item[7.] The preamble is sent over the \ac{PRACH}. The interleaving pattern selected in Step 1 determines the \acp{PO} where the individual segments must be transmitted (Figure \ref{fig:SBIDMA_detail}(f)).
\end{itemize}

\subsubsection{Receiver Chain}

\begin{figure}
    \centering
    \includegraphics[width = 0.9\textwidth]{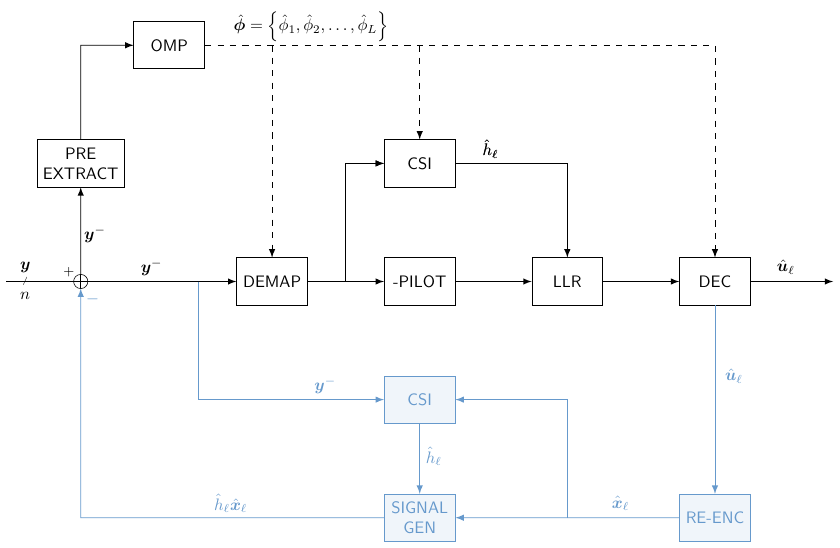}
    \caption{Schematic description of the \ac{SB-IDMA} receiver chain.}
    \label{fig:SBIDMA_DEC}
\end{figure}

The receiver behavior is largely based on the procedures described in Box 3, and already adopted for the \ac{2SRA} simulations. Iterative \ac{TIN}-\ac{SIC} decoding is assumed, as visualized in Figure \ref{fig:SBIDMA_DEC}. Some important details about the receiver behavior follow:
\begin{itemize}
    \item In each iteration of \ac{TIN}-\ac{SIC}, as set of $L$ preambles is obtained via \ac{OMP}, applied to the observation of the \ac{PRACH}. 
    \item For each detected preamble, the sequence of \acp{PO} used to transmit the user segments is determined, as well as the sequence of pilot fields.
    \item In each \ac{PO}, channel estimation and interference-plus-noise power estimation are performed (see Box 3). \acp{LLR} for the codeword bits are computed accordingly. The $d$ \acp{LLR} associated with the repetition of each codeword bit are combined (summed) and passed to the decoder of $\code$. 
    \item Assuming an incomplete decoding algorithm, the outcome of decoding can be either a detected error or a valid codeword decision. In the former case, no further action is needed: the decoder output is simply discarded. In the latter case, an additional error detection step is performed by (a) computing the hash of the decoded message and (b) comparing it with the preamble index associated with the decoding attempt: if the two indexes are different, the decoder output is discarded. Otherwise, the decoded message is deemed to be correct, and it is passed at the receiver output.
    \item For decoded messages that are considered correct, \ac{SIC} is performed. In particular, as for the discussion that follows Box 3, the channel coefficient is re-estimated, for each segment, using the decoded data as an extended pilot field. The interference contribution of each segment is then removed from the respective \ac{PO}. Similarly, the interference contribution of the corresponding preamble is cancelled from the \ac{PRACH} observation, with channel coefficient provided by the \ac{OMP} algorithm.
\end{itemize}

\subsubsection{SB-IDMA: Performance}

In the tested \ac{SB-IDMA} configurations, two types of error correcting codes are considered: the \ac{5GNR} \ac{LDPC} codes, and the \ac{5GNR} polar codes. In the former case, \ac{BP} decoding is assumed with $50$ decoding iterations. In the latter case, adaptive \ac{SCL} decoding is used \cite{Li2012}, with a maximum list size set to $128$.

\textit{\textsf{Gaussian MAC Performance}}

A first set of results is provided for the Gaussian \ac{MAC} in Figure \ref{fig:AWGN}. The configurations have been tailored for $n=15000$ \acp{c.c.u.} ($30000$ real \acp{c.u.}), hence are directly comparable with the schemes whose performance is provided on the same chart. The details of the \ac{SB-IDMA} configurations are summarized in Table \ref{tab:SBIDMA_GMAC}. The configurations are somehow ``extreme'', in the sense that they employ \acp{PO} of relatively small size ($25$ complex channel uses each). As a result, each transmission yields a relatively large number of segments ($n_s = 80$). Preambles and pilots have been drawn randomly, with symbols that are independently distributed according to a circularly-symmetric Gaussian distribution (see Box 1).

The performance of \ac{SB-IDMA} with the $(500,100)$ \ac{LDPC} code (\ref{fig:AWGN:SBIDMA_LDPC}) from the \ac{NR} standard is with $1.8$ dB from the achievability bound (\ref{fig:AWGN:RCU}) up to a moderate-large number of active users, competing with some of the best schemes from literature. The gap from the bound can be dissected in two main components. A $0.6$ dB loss is caused by the energy overhead introduced by the preamble. The remaining $1.2$ dB loss can be attributed to the \ac{LDPC} code: that is the gap displayed by the $(500,100)$ \ac{5GNR} \ac{LDPC} code, at a block error rate of $5 \times 10^{-2}$, with respect to the \ac{RCU} bound of \cite{PPV10} over the single-user \ac{AWGN} channel. The configuration that employs the \ac{5GNR} polar code (\ref{fig:AWGN:SBIDMA_POLAR}) exploits the fact that the polar code rate can be further lowered, with respect to the \ac{LDPC} code. A rate-$1/10$ polar code is hence used. The performance over the overall scheme results in a gain of about $0.8$ dB over its \ac{LDPC}-based counterpart, operating within $1$ dB from the \ac{RCU} up to $80$ active users.

\textit{\textsf{Quasi-Static Fading MAC Performance}}

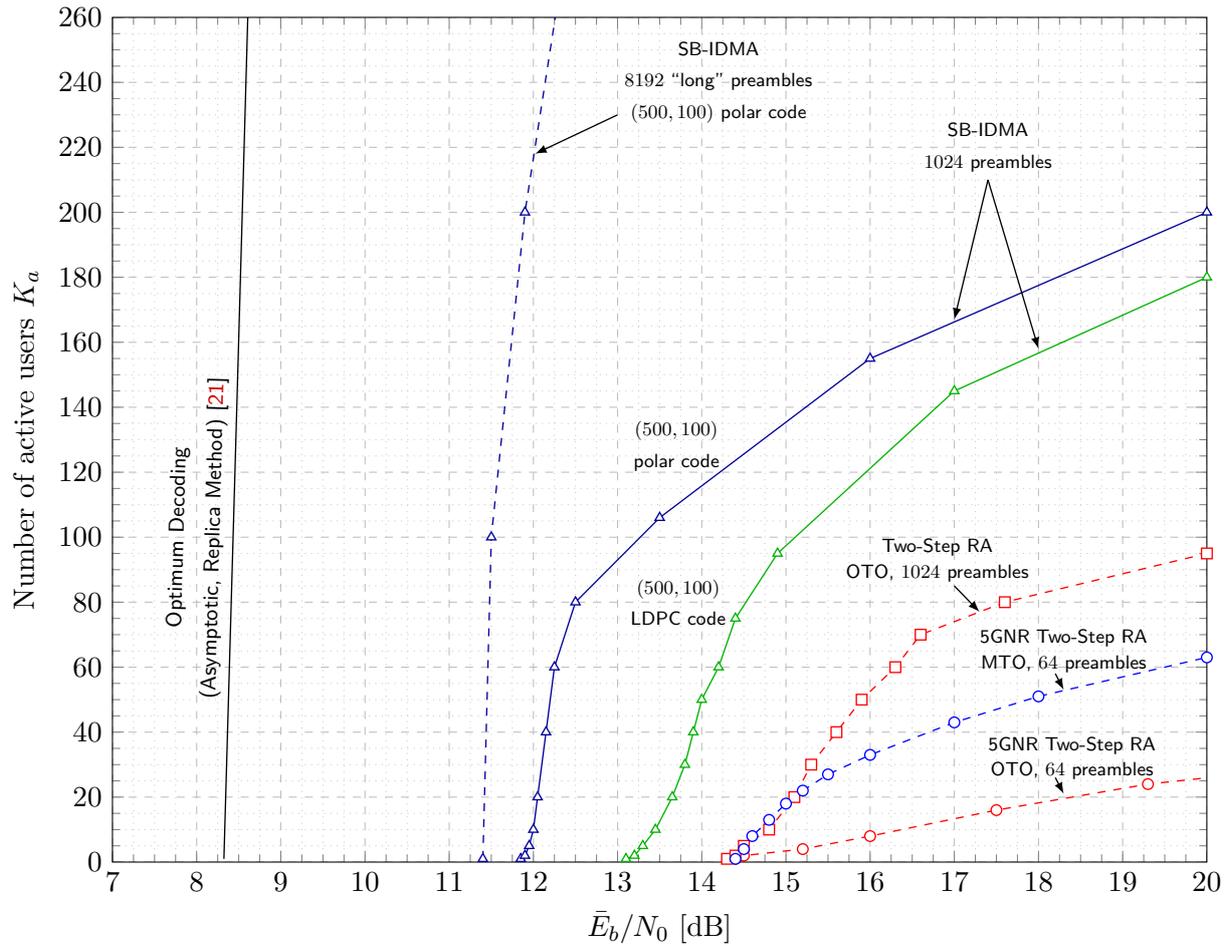
\begin{figure}
    \centering
    \begin{tikzpicture}
	\pgfplotsset{set layers,compat=1.3}
	
	\begin{axis}[%
		width=0.9\textwidth,
		height=0.7\textwidth,
		scale only axis,
		ymin=0,
		ymax=260,
		ylabel={Number of active users $K_a$},
		xmin=7,
		xmax=20,
		xlabel={$\bar{E}_b/N_0$ [dB]},
		xlabel near ticks,
		grid=both,
		grid style={dashed,gray!50},
		minor grid style={dotted,gray!40},
		minor tick num=3,
		legend cell align={left},
		legend style={font= \footnotesize, draw = none},
		]

		\addplot [color=red,mark=*,line width=0.5pt,dashed,mark options={solid,fill=white}]
		table[y=K,x=snr] {charts/TwoStep500LDPC_fading_2xPre.txt};\label{fig:SISO2:2SRA_OTO}
		
		\addplot [color=red,mark=square*,line width=0.5pt,dashed,mark options={solid,fill=white}]
		table[y=K,x=snr] {charts/TwoStep500LDPC_fading_1024preambles_2xPre.txt};\label{fig:SISO2:2SRA_OTO1024}

		\addplot [color=blue,mark=*,line width=0.5pt,dashed,mark options={solid,fill=white}]
		table[y=K,x=snr] {charts/TwoStep_Fading_Alternate_Config.txt};\label{fig:SISO2:2SRA_MTO}
		
		\node[align=center,scale = 0.8] at (axis cs: 18.4,32) {\footnotesize \textsf{5GNR Two-Step RA} \\[-1mm] \footnotesize \textsf{OTO, $64$ preambles}};
		\draw[-latex,line width = 0.5pt] (axis cs: 18.2,26) -- (axis cs: 18.3,21);
		
		\node[align=center,scale = 0.8] at (axis cs: 18.3,65) {\footnotesize \textsf{5GNR Two-Step RA} \\[-1mm] \footnotesize \textsf{MTO, $64$ preambles}};
		\draw[-latex,line width = 0.5pt] (axis cs: 18.2,58) -- (axis cs: 18.3,53);
		
		\node[align=center,scale = 0.8] at (axis cs: 16.8,93) {\footnotesize \textsf{Two-Step RA} \\[-1mm] \footnotesize \textsf{OTO, $1024$ preambles}};
		\draw[-latex,line width = 0.5pt] (axis cs: 17,85) -- (axis cs: 17.3,77);
			
		\addplot [color=black,line width=0.5pt]
		table[y=K,x=snr] {charts/fading_replica.txt};\label{fig:SISO2:BOUND}
		
		\node[scale = 0.9,align = center,color = black, rotate  = 88.5] at (axis cs: 8.0,100) {\footnotesize \textsf{Optimum Decoding} \\ \footnotesize \textsf{(Asymptotic, Replica Method) \cite{Frolov20}} };
		
		\addplot [color=green!70!black,mark=triangle*,line width=0.5pt,mark options={solid,fill=white}]
		table[y=K,x=snr] {charts/SBIDMA500LDPCd3_fading_1024preambles_2xPre.txt};\label{fig:SISO2:SBIDMA_LDPC}
		
		\addplot [color=blue!60!black,mark=triangle*,line width=0.5pt,mark options={solid,fill=white}]
		table[y=K,x=snr] {charts/SBIDMA500polard3_fading_1024preambles_2xPre.txt};\label{fig:SISO2:SBIDMA_POLAR}
		
		\node[align=center, scale = 0.8] at (axis cs: 17.4,220) {\footnotesize \textsf{SB-IDMA} \\ \footnotesize \textsf{$1024$ preambles}};
		\draw[-latex,line width = 0.5pt] (axis cs: 17.4,210) -- (axis cs: 17,167);
		\draw[-latex,line width = 0.5pt] (axis cs: 17.4,210) -- (axis cs: 18,158);
		
		\node[align=center, scale = 0.8] at (axis cs: 13.7,128) {\footnotesize $(500,100)$ \\ \footnotesize \textsf{polar code}};
		
		\node[align=center, scale = 0.8] at (axis cs: 13.72,80) {\footnotesize $(500,100)$ \\ \footnotesize \textsf{LDPC code}};
		
		\addplot [color=blue!60!black,dashed,mark=triangle*,line width=0.5pt,mark options={solid,fill=white},forget plot]
		table[y=K,x=snr] {charts/SBIDMA500Polard3_fading_8192preambles_1778xPre1.12.txt};\label{fig:SISO2:POLARSUPER}
		
		\node[align=center, scale = 0.8] at (axis cs: 14.2,240) {\footnotesize \textsf{SB-IDMA} \\ \footnotesize \textsf{$8192$ ``long'' preambles}\\ \footnotesize \textsf{$(500,100)$ polar code}};
		\draw[-latex,line width = 0.5pt] (axis cs: 13,230) -- (axis cs: 12.02,218);
		
	\end{axis}
\end{tikzpicture}
    \caption{Number of supported active users vs. average \ac{SNR} for a target $\PUPE = 10^{-1}$. Quasi-static Rayleigh fading channel, $n\approx 20000$ channel uses. Single antenna at the base station.}
    \label{fig:fading_SISO2}
\end{figure}

A second set of results was obtained over the quasi-static fading \ac{MAC} (Figure \ref{fig:fading_SISO2}). The parameters used for the simulations are given in Table \ref{tab:SBIDMA_FMAC}, and are closely aligned to the \ac{2SRA} \ac{OTO} configuration of Table \ref{tab:2SRA}. The performance of the \ac{LDPC}-based (\ref{fig:SISO2:SBIDMA_LDPC}) and of the polar-based (\ref{fig:SISO2:SBIDMA_POLAR}) scheme shows a remarkable gain over the various \ac{2SRA} configurations (including the non-standard extension using $1024$ preambles). A saturation of the number of supported users, that happens around $160 \div 200$ users in the polar code case, can still be observed, with a two-fold improvement over the \ac{2SRA} \ac{OTO} mapping with $1024$ preambles (\ref{fig:SISO2:2SRA_OTO1024}). The saturation is here caused by the limited number of preambles, and can be largely improved by enabling a even larger preamble set. However, as observed for the \ac{2SRA} case, the number of channel uses allocated to the \ac{PRACH} ($278$) limits the number of preambles that can be supported under \ac{OMP} detection. To enlarge the preamble set, longer preambles can be used. On the same chart, the performance of an \ac{SB-IDMA} configuration that trades the number of \acp{PO} ($N=59$)
for longer preambles ($12\times 139$ \acp{c.c.u.}) is shown (\ref{fig:SISO2:POLARSUPER}). To limit the energy overhead caused by the longer preamble, a power back-off of $10$ dB is applied to the preambles.
The resulting performance allows to support more than $260$ active users, without displaying signs of saturation.

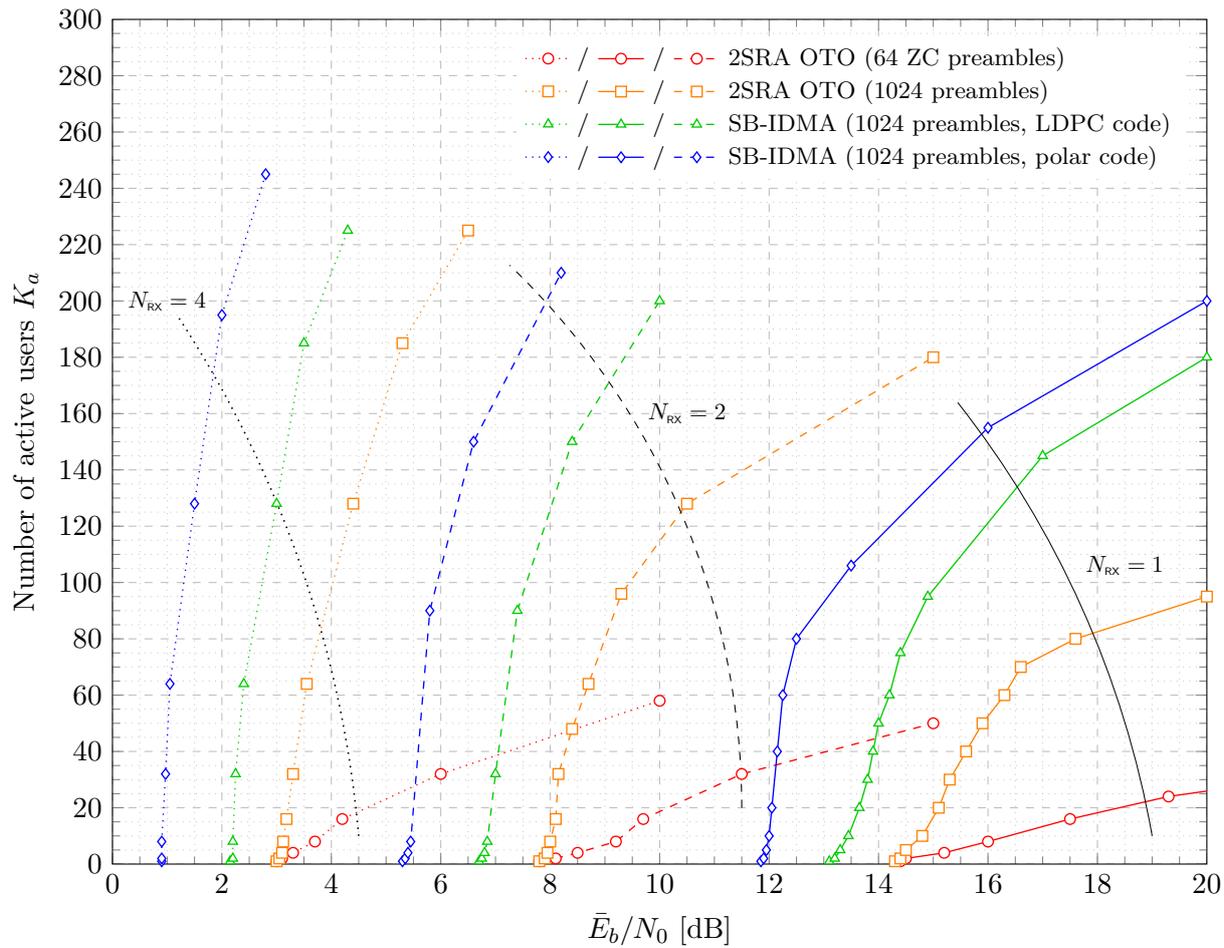
\begin{figure}
    \centering
    \begin{tikzpicture}
	\pgfplotsset{set layers,compat=1.3}
	
	\begin{axis}[%
		width=0.9\textwidth,
		height=0.7\textwidth,
		scale only axis,
		ymin=0,
		ymax=300,
		ylabel={Number of active users $K_a$},
		xmin=0,
		xmax=20,
		xlabel={$\bar{E}_b/N_0$ [dB]},
		xlabel near ticks,
		grid=both,
		grid style={dashed,gray!50},
		minor grid style={dotted,gray!40},
		minor tick num=3,
		legend cell align={left},
		legend style={font= \footnotesize, draw = none},
		]

\addplot [color=red,mark=*,line width=0.5pt,mark options={solid,fill=white},forget plot]
table[y=K,x=snr] {charts/TwoStep500LDPC_fading_2xPre.txt};\label{fig:SIMO:2SRA:A1}
\addplot [color=red,dashed,mark=*,line width=0.5pt,mark options={solid,fill=white},forget plot]
table[y=K,x=snr] {charts/TwoStep500LDPC_fading_2xPre_A2.txt};\label{fig:SIMO:2SRA:A2}
\addplot [color=red,dotted,mark=*,line width=0.5pt,mark options={solid,fill=white}]
table[y=K,x=snr] {charts/TwoStep500LDPC_fading_2xPre_A4.txt};\label{fig:SIMO:2SRA:A4}
\addlegendentry{/ \raisebox{-0.1mm}{\ref{fig:SIMO:2SRA:A1}} / \raisebox{-0.1mm}{\ref{fig:SIMO:2SRA:A2}} 2SRA OTO ($64$ ZC preambles)};

\addplot [color=orange,mark=square*,line width=0.5pt,mark options={solid,fill=white},forget plot]
table[y=K,x=snr] {charts/TwoStep500LDPC_fading_1024preambles_2xPre.txt};\label{fig:SIMO:2SRA1024:A1}
\addplot [color=orange,dashed,mark=square*,line width=0.5pt,mark options={solid,fill=white},forget plot]
table[y=K,x=snr] {charts/TwoStep500LDPC_fading_1024preambles_2xPre_A2.txt};\label{fig:SIMO:2SRA1024:A2}
\addplot [color=orange,dotted,mark=square*,line width=0.5pt,mark options={solid,fill=white}]
table[y=K,x=snr] {charts/TwoStep500LDPC_fading_1024preambles_2xPre_A4.txt};\label{fig:SIMO:2SRA1024:A4}
\addlegendentry{/ \raisebox{-0.1mm}{\ref{fig:SIMO:2SRA1024:A1}} / \raisebox{-0.1mm}{\ref{fig:SIMO:2SRA1024:A2}} 2SRA OTO ($1024$ preambles)};

\addplot [color=green!80!black,mark=triangle*,line width=0.5pt,mark options={solid,fill=white},forget plot]
table[y=K,x=snr] {charts/SBIDMA500LDPCd3_fading_1024preambles_2xPre.txt};\label{fig:SIMO:LDPC:A1}
\addplot [color=green!80!black,dashed,mark=triangle*,line width=0.5pt,mark options={solid,fill=white},forget plot]
table[y=K,x=snr] {charts/SBIDMA500LDPCd3_fading_1024preambles_2xPre_A2.txt};\label{fig:SIMO:LDPC:A2}
\addplot [color=green!80!black,dotted,mark=triangle*,line width=0.5pt,mark options={solid,fill=white}]
table[y=K,x=snr] {charts/SBIDMA500LDPCd3_fading_1024preambles_2xPre_A4.txt};\label{fig:SIMO:LDPC:A4}
\addlegendentry{/ \raisebox{-0.1mm}{\ref{fig:SIMO:LDPC:A1}} / \raisebox{-0.1mm}{\ref{fig:SIMO:LDPC:A2}} SB-IDMA ($1024$ preambles, LDPC code)};

\addplot [color=blue,mark=diamond*,line width=0.5pt,mark options={solid,fill=white},forget plot]
table[y=K,x=snr] {charts/SBIDMA500polard3_fading_1024preambles_2xPre.txt};\label{fig:SIMO:POLAR:A1}
\addplot [color=blue,dashed,mark=diamond*,line width=0.5pt,mark options={solid,fill=white},forget plot]
table[y=K,x=snr] {charts/SBIDMA500polard3_fading_1024preambles_2xPre_A2.txt};\label{fig:SIMO:POLAR:A2}
\addplot [color=blue,dotted,mark=diamond*,line width=0.5pt,mark options={solid,fill=white}]
table[y=K,x=snr] {charts/SBIDMA500polard3_fading_1024preambles_2xPre_A4.txt};\label{fig:SIMO:POLAR:A4}
\addlegendentry{/ \raisebox{-0.1mm}{\ref{fig:SIMO:POLAR:A1}} / \raisebox{-0.1mm}{\ref{fig:SIMO:POLAR:A2}} SB-IDMA ($1024$ preambles, polar code)};

\draw[line width=0.6pt,dotted] (45,10) arc (5:33:15cm);
\node[align = center,scale = 0.8] at (axis cs: 1,200) {$\NRX=4$};

\draw[line width=0.4pt,dashed] (115,20) arc (0:46:10cm);
\node[align = center,scale = 0.8] at (axis cs: 10.5,160) {$\NRX=2$};		

\draw[line width=0.3pt] (190,10) arc (10:38:13cm);
\node[align = center,scale = 0.8] at (axis cs: 18.5,106) {$\NRX=1$};		
		
	\end{axis}
\end{tikzpicture}
    \caption{Number of supported active users vs. average \ac{SNR} for a target $\PUPE = 10^{-1}$. Quasi-static Rayleigh fading channel, $n\approx 20000$ channel uses. Multiple antennas at the base station.}
    \label{fig:fading_SIMO}
\end{figure}

A final set of results, aiming at assessing the effect of multiple antennas at the base station, is provided in Figure \ref{fig:fading_SIMO}. Here, the number of receiver antennas is moderate, and set to $\NRX = 2$ and to $\NRX = 4$. The configurations used for the simulations are the ones described in Table \ref{tab:2SRA} and in Table \ref{tab:SBIDMA_FMAC}. The results confirm all the insights gathered in the previous simulation setups.

\begin{table*}
\caption{Parameters used for \ac{SB-IDMA} simulations -- Gaussian \ac{MAC}.}
\centering
{\tabulinesep=1mm
\begin{tabu}{lccc}
\hline
\hline
Parameter & \ac{LDPC}-based & Polar-based & Unit\\
\hline
Information bits per user & $100$ & $100$ & bits\\
Frame length ($n$) & $15000$ & $15000$ & c.c.u.\\
\# \acp{PO} ($N$)    & $589$  & $589$ & - \\
\ac{PO} size ($\POlength$)    & $25$  & $25$ & c.c.u. \\
Preamble length     & $275$ & $275$ & c.c.u.\\
\# Preambles & $2048$ & $2048$ & - \\
Channel code & 5GNR LDPC (BG2) & 5GNR Polar (CRC-$11$) & - \\
Block length ($n_c$) & $500$ & $1000$ & bits \\
Modulation & QPSK & QPSK & - \\
Repetition rate ($d$) & $8$ & $4$ & - \\
\# Segments ($n_s$) & $80$ & $80$ & - \\
\hline
\hline
\end{tabu}}
\label{tab:SBIDMA_GMAC}
\end{table*}

\begin{table*}
\caption{Parameters used for \ac{SB-IDMA} simulations -- Quasi-static fading \ac{MAC}.}
\centering
{\tabulinesep=1mm
\begin{tabu}{lccc}
\hline
\hline
Parameter & \ac{LDPC}-based & Polar-based & Unit\\
\hline
Information bits per user & $100$ & $100$ & bits\\
Frame length ($n$) & $19478$ & $19478$ & c.c.u.\\
\# \acp{PO} ($N$)    & $64$  & $64$ & - \\
\ac{PO} size ($\POlength$)    & $300$  & $300$ & c.c.u. \\
Preamble length     & $278$ & $278$ & c.c.u.\\
\# Preambles & $1024$ & $1024$ & - \\
Channel code & 5GNR LDPC (BG2) & 5GNR Polar (CRC-$11$) & - \\
Block length ($n_c$) & $500$ & $500$ & bits \\
Modulation & QPSK & QPSK & - \\
Repetition rate ($d$) & $3$ & $3$ & - \\
\# Segments ($n_s$) & $3$ & $3$ & - \\
\# pilot symbols per \ac{PO} & $50$ & $50$ & - \\
\hline
\hline
\end{tabu}}
\label{tab:SBIDMA_FMAC}
\end{table*}

\clearpage

\clearpage

\clearpage
\section{Conclusions}\label{sec:conclusions}

This report provided a first investigation on possible improvements of the random access channel in future evolutions of the 3GPP standards. The analysis addressed the design of massive random access schemes for grant-free access, taking as reference the initial component of the \ac{2SRA} protocol included in the \ac{5GNR} specification. The study identified three directions that may yield substantial improvements to the existing protocol, namely
\begin{enumerate}
	\item The extension of the set of preambles to be used over the \ac{PRACH}, together with the introduction of preamble-dependent pilot fields;
	\item The definition of a more flexible use of \acp{PO}, e.g., by allowing transmission to take place over several \acp{PO}.
	\item The adoption of polar codes, when short data packets have to be transmitted.
\end{enumerate}
The identified improvement points have been embodied in a new scheme -- \ac{SB-IDMA} --  which requires relatively minor updates of the specification, while delivering sizable gains over both Gaussian and fading \acp{MAC}.

\clearpage


\clearpage
\appendix

\section*{Acronyms and Notation}

\subsection*{List of Acronyms}
\begin{acronym}[XXXXXXXX]
        \acro{2SRA}{two-step random access}
        \acro{4SRA}{four-step random access}
        \acro{5GNR}{5G New Radio}
        \acro{AWGN}{additive white Gaussian noise}
        \acro{BP}{belief propagation}
        \acro{BS}{base station}
        \acro{CCS}{coded compressed sensing}
        \acro{c.c.u.}{complex channel use}
        \acro{CP}{cyclic prefix}
        \acro{CRDSA}{contention resolution diversity slotted Aloha}
        \acro{CS}{compressed sensing}
        \acro{CSA}{coded slotted Aloha}
        \acro{c.u.}{channel use}
        \acro{IDMA}{interleaver division multiple access}
        \acro{IoT}{Internet of Things}
        \acro{IRSA}{irregular repetition slotted Aloha}
        \acro{ISI}{inter-symbol interference}
	    \acro{LDPC}{low-density parity-check}
        \acro{LLR}{log-likelihood ratio}
        \acro{LTE}{Long Term Evolution}
        \acro{MAC}{multiple access}
        \acro{MMSE}{minimum mean square error}
        \acro{MPR}{multi-packet reception}
        \acro{MTC}{machine-type communication}
        \acro{MTO}{many-to-one}
        \acro{NB-IoT}{Narrowband IoT}
        \acro{NR}{new radio}
        \acro{OFDM}{orthogonal frequency-division modulation}
        \acro{OMP}{orthogonal matching pursuit}
        \acro{OTO}{one-to-one}
        \acro{PAM}{pulse-amplitude modulation}
        \acro{PBCH}{physical broadcast channel}
        \acro{PDSCH}{physical downlink shared channel}
        \acro{PO}{PUSCH occasion}
        \acro{PRACH}{physical random access channel}
        \acro{PRB}{physical resource block}
        \acro{PSS}{primary synchronization signal}
        \acro{PUPE}{per-user probability of error}
        \acro{PUSCH}{physical uplink shared channel}
        \acro{QPSK}{quadrature phase shift keying}
        \acro{RA}{random access}
        \acro{RCU}{random coding union}
        \acro{SB-IDMA}{sparse block interleaver division multiple access}
        \acro{SCL}{successive cancellation list}
        \acro{SCS}{sub-carrier spacing}
        \acro{SIC}{successive interference cancellation}
        \acro{SNR}{signal-to-noise ratio}
        \acro{SPARC}{sparse regression code}
        \acro{SSS}{secondary synchronization signal}
        \acro{TA}{time advance}
        \acro{TIN}{treat-interference-as-noise}
        \acro{UT}{user terminal}
        \acro{UMAC}{unsourced multiple access}
\end{acronym}
\clearpage

\subsection*{Notation Summary}
\begin{tabular}{@{}p{2.4cm}l@{}}
$\code$ & Error correcting code\\
$E_b$ & Energy per information bit\\
$k$ & Number of information bits per active user\\
$K$ & Total number of users\\
$K_a$ & Number of active users\\
$n$ & Frame length\\
$n_s$ & Number of segments (SB-IDMA)\\
$N$ & Number of PUSCH occasions\\
$N_0$ & Single-sided noise power spectral density\\
$n_c$ & Block length of the error correcting code\\
$\POlength$ & number of channel uses in a \ac{PO}\\
$\preamblelength$ & Preamble length in channel uses\\
$\NRX$ & Number of antennas at the base station\\
$\PUPE$ & Per-user probability of error\\
\end{tabular}

\end{document}